\documentclass[aps,preprint,showpacs]{revtex4}
\usepackage{latexsym}
\usepackage{graphicx}
\usepackage{epsfig}
\usepackage{dcolumn}
\usepackage{natbib}
\usepackage{amsmath}
\usepackage{amssymb}

\begin{document}

\title{Line creep in paper peeling}

\author{Jari Rosti}
\affiliation{Department of Engineering Physics, Helsinki University of Technology, 
             FIN-02015 HUT, email: firstname.secondname@tkk.fi}

\author{Juha Koivisto}
\affiliation{Department of Engineering Physics, Helsinki University of Technology, 
             FIN-02015 HUT, email: firstname.secondname@tkk.fi}

\author{Paola Traversa}
\affiliation{Universit\'e Joseph Fourier, CNRS, LGIT, BP 53, 38041 Grenoble, France}

\author{Xavier Illa}
\affiliation{Department of Engineering Physics, Helsinki University of Technology, 
             FIN-02015 HUT, email: firstname.secondname@tkk.fi}

\author{Jean-Robert Grasso}
\affiliation{Universit\'e Joseph Fourier, CNRS, LGIT, BP 53, 38041 Grenoble, France}

\author{Mikko J. Alava}
\affiliation{Department of Engineering Physics, Helsinki University of Technology, 
             FIN-02015 HUT, email: firstname.secondname@tkk.fi}

\begin{abstract}
The dynamics of a "peeling front" or an elastic line is studied
under creep (constant load) conditions. Our experiments show 
an exponential dependence of the creep velocity on the inverse
force (mass) applied. In particular, the dynamical correlations of
the avalanche activity are discussed here. We compare various
avalanche statistics to those of a line depinning model with non-local elasticity,
and study various measures of the experimental avalanche-avalanche and temporal
correlations such as the autocorrelation function of the released
energy and aftershock activity. From all these we
conclude, that internal avalanche dynamics seems to follow "line
depinning" -like behavior, in rough agreement with the depinning
model. Meanwhile, the correlations reveal subtle complications not
implied by depinning theory. Moreover, we also show how these results can 
be understood from a geophysical point of view.
\end{abstract}

\maketitle
\date{\today}

\section{Introduction}
Creep is one of the fascinating topics in fracture for a physicist:
the deformation and final fracture of a sample follow empirical laws
with a rich phenomenology. It is expected that there are
similarities and differences with "static" fracture encountered in
brittle materials such that so-called "time-dependent rheology" is
not relevant \cite{advphys}. However, the phenomenon of creep is
visible in most any setting regardless of whatever a tensile test
might indicate about the typical material response. A particular
scenario where one can study creep is the advancement of a single
crack under a constant driving force. One can study this in simple
paper sheets, and for quite some time it has been noticed that this
involves statistical phenomena, an intermittent response which could
be characterized by "avalanches", in particular of acoustic emission
(AE) events \cite{sethrev,Kertesz,oma,santucci}.

A particular experiment we analyze in this work is related to the
dynamics of a {\em crack line} as it moves through a sample, largely
constrained on a plane. This can be achieved in the case of paper in
the so-called Peel-In-Nip (PIN) geometry (see below for a description).
The tensile case has been already reported in Ref. \cite{epl} and an early
account of the creep results published as Ref. \cite{koivisto}. The mathematical
description of the line is a crack position $h(x,t)$,  where $h$ is the position
coordinate along the direction of line propagation and $x$ is the 
coordinate perpendicular to $h$. On the average, 
the crack moves with the {\em creep velocity} $v$ ($\bar{h} = vt$).

The problem has here, as in other such examples (the Oslo plexiglass
experiment \cite{maloy1,maloy2}), three important ingredients: randomness in that
the peeling line experiences a disordered environment coming from
the fiber network structure, a driving force $K_{eff}$ or a stress
intensity factor, and the self-coupling of the interfacial profile
$h$. In this particular problem, it takes place via a long-range
elastic kernel \cite{Fisher}, expected to scale as $1/x$ or as $k$
in Fourier space.

For a constant force $K_{eff}$ the dynamics exhibits a depinning
transition, of non-equilibrium statistical mechanics. This implies a
phase diagram for $v(K_{eff})$. The crack begins to move ($v>0$) at a
critical value $K_c$ of $K_{eff}$ such that for $K_{eff} > K_c$. In
the proximity of $K_c$ the line geometry is a self-affine fractal
with a roughness exponent $\zeta$. The planar crack problem
\cite{Fisher2,schmitt}  has been studied theoretically via
renormalization group calculations and numerical simulations, and
via other experiments as noted above. The roughness exponent of
theory $\zeta_{theory} \sim 0.39$ has traditionally been considered
to be absent from experiments \cite{Rosso2002,maloy1,maloy2,krauth}, but recent
results of Santucci et al. imply that the regime might be visible
upon coarse-graining. Imaging experiments prove in that case that as
expected the line moves in avalanches, and the avalanche size $s$
distribution seems to have the form $P(s) \sim s^{-1.6 \dots -1.7}$
\cite{maloy1,maloy2}.

Here we look at the scenario of creep for the PIN geometry. This
subject is such that ordinary "fracture creep" and the particular
scenario related to depinning transitions coincide.  The creep of
elastic lines becomes important for $K_{eff} \leq K_c$ since
thermally assisted movement due to fluctuations takes place with a
non-zero temperature \cite{creep,chauve,kolton}. In usual
depinning, it is assumed that thermal fluctuations nucleate
``avalanches' which derive their properties from zero-temperature
depinning, and the avalanches then translate into a finite velocity
$v_{creep}>0$. There are two interesting differences in the fracture
line creep to other such in depinning. First, the line elasticity is
non-local, and second, in materials (such as paper here) where there
is no healing, the line motion is irreversible, there are no
fluctuations in metastable states as in the case of magnetic domain
walls, for instance.

In this scenario, the creep velocity becomes a function of the
applied stress intensity factor and the temperature, $v_{creep}=
v_{creep} (K_{eff},T)$.  As creep takes place via nucleation events
over energy barriers \cite{creep}, the description of those barriers
is of fundamental importance.  One can show by scaling arguments and
more refined renormalization group treatments that the outcome has
the form of the following creep formula
\begin{equation}
v_{creep} \sim \exp{(-C/K_{eff}^\mu)} \label{vcreep}.
\end{equation}
This gives the relation to the driving force $K_{eff}$ using the
creep exponent, $\mu$. The value of the exponent depends on the
elastic interactions and the dimension of the moving object (a
line), and we expect
\begin{equation}
\mu = \theta/\nu = \frac{1-\alpha + 2\zeta}{\alpha-\zeta}.
\label{expo}
\end{equation}
The exponents $\theta$, $\nu$, and $\zeta$ denote the energy
fluctuation, correlation length, and {\em equilibrium} roughness
exponents. All these exponents are functions of $\alpha$, the
$k$-space decay exponent of the elastic kernel. For long range
elasticity, one would assume $\alpha=1$.

\begin{figure}[htb]
\begin{center}
\epsfig{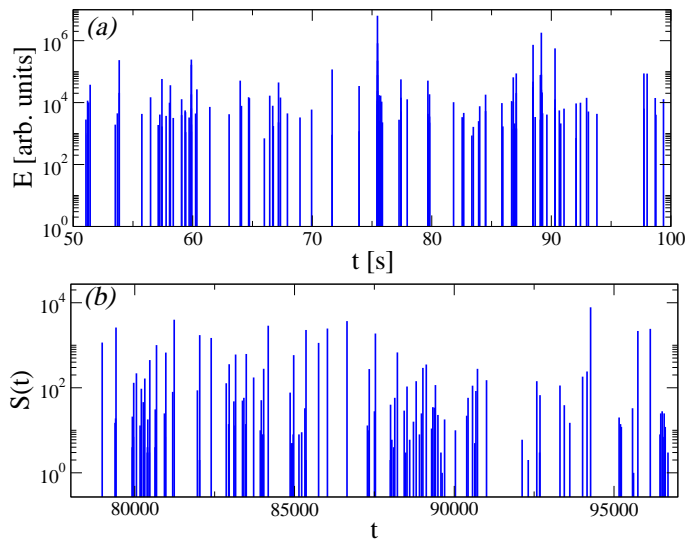}
\end{center}
\caption{\label{fig:et} Activity as a function of time inside a given time window
(a) for the creep experiment with 410g load, and (b) for simulations with $f=1.87$ and $T_p=0.002$.
In both cases we neglect the duration of the avalanche and we only take into acount the starting time and the size of each avalanche, obtaining a data series $\{t_i,E_i\}$ for the experiments and $\{t_i,S_i\}$ for simulations (definition of $S_i$ is given in Section \ref{sims})}
\end{figure}

The fundamental formula of Eq.~(\ref{expo}) has been confirmed in
the particular case of 1+1-dimensional domain walls and other
experiments \cite{lemerle,crex}. We have ourselves reported on
results, which show an inverse exponential dependence of $v_{creep}(m)
\sim \exp{(-1/m)}$, where $m$ is the applied mass in the experiment
(see below), as is appropriate for non-local line elasticity with an
{\em equilibrium roughness exponent} of $\zeta=1/3$. In the current
work we go further by two important steps. First, we consider creep
simulations of an appropriate non-local line model and compare the
avalanche statistics and $v(m)$ to those from the experiments (see
Fig. \ref{fig:et} for an example of the activity timeseries from an
experiment and a simulation). Then, we ask the fundamental question:
what can be stated of the correlations? This relates to the
timeseries of released energy, to aftershock rates
and we present extensive evidence. The experimental signatures
show subtle correlations that are rather different from what one would
expect from the (depinning) creep problem with non-existing avalanche to
avalanche correlations.

The structure of the rest of the paper is as follows. In the next
Section, we discuss the experimental setup and the simulation model.
Section III shows results on $v(m)$ both from experiment and
simulation. In Section IV we present data on avalanche statistics
again comparing the two cases. Section V offers an extensive
analysis of correlations by using a number of techniques to look at
the experiment. Finally, Section VI finishes with
Conclusions and a Discussion.

\section{Methods}

\subsection{Experiment}

\begin{figure}[htb]
\begin{center}
\epsfig{file=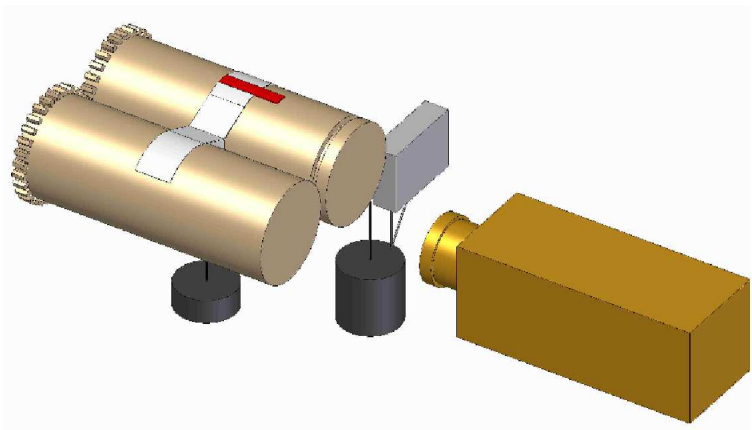,width=10.cm,clip=}
\end{center}
\caption{\label{fig:rullat} Experimental setup for peeling experiment. The paper (white) is peeled
between two cylinders (copper) separated by a few millimeters. The
driving force is generated by a larger hanging weight (black). A
smaller weight adjusts the peeling angle.
The AE and distance data are collected by piezo transducer (red) and a
laser sensor (gray).}
\end{figure}

In Figure \ref{fig:rullat} we show the apparatus \cite{epl}.
The failure line can be located along the ridge, in the
center of the the Y-shaped construction formed by the
unpeeled part of the sheet (below) and the two parts
separated by the advancing line. Diagnostics consist
of an Omron Z4D-F04 laser distance sensor for
the displacement, and a standard plate-like piezoelectric sensor \cite{epl}.
It is attached to the
setup inside one of the rolls visible in Fig.~\ref{fig:rullat},
and the signal is filtered
and amplified using standard techniques. The data
aquisition card gives us four channels at 312.5 kHz
per channel. We finally threshold the AE data. The
displacement data is as expected highly correlated with
the corresponding AE, but the latter turns out to include much
less noise and thus convenient to study.
For paper, we use perfectly standard copy paper,
with an areal mass or basis weight of 80 g/m$^2$.
Industrial paper has two principal directions, called
the ``Cross'' and ``Machine'' Directions (CD/MD).
The deformation characteristics are much more ductile
in CD than in MD, but the fracture stress is higher
in MD \cite{Alava2006}. We tested a number of samples for both directions,
with strips of width 30 mm. The weight used for the
creep ranges from 380 g to 450 g for CD case and from 450 g to 533 g for MD case.
The mechanical (and creep) properties of paper depend
on the temperature and humidity. In our setup
both remain at constant levels during experiments,
and the typical pair values for  environment is
40 rH and 26 $^o$C.

\subsection{Simulations}
\label{sims}

We want to simulate the evolution of a discrete long-range elastic
line of size $L$ in a disordered media. The line is characterized by a vector
of integer heights $\left\{ h_1, \ldots h_L \right \}$ with periodic
boundary conditions, although the experiment does not present periodic bound conditions.

The long-range elastic force \cite{Tanguy1998} acting on a string
element is given by
\begin{equation}
f^{elastic}_i =k_o \left( \frac{\pi}{L} \right )^2
    \sum_{\substack{j=1 \\ j \neq i }}^{L} \frac{h_j-h_i}{\sin \left( {\frac{x_j-x_i}{L}\pi} \right)},
\end{equation}
where all forces on all sites can be computed in a $ L \log L $
operations using a fast-Fourier-transform (FFT) algorithm
\cite{Duemmer2007}. Simulations are done using $k_o=0.01$ and
$L=1024$.
The random force due to the quenched disorder may be obtained
from a standard normal distribution, i.e a Gaussian distribution
with zero mean and a variance of one $f^{random}_i=N(0,1)$.
Then, the total force acting in a given element of the string is
$f_i= f^{elastic}_i + f^{random}_i + f$, where $f$ is the external
applied force.

At this point, we need to introduce a dynamics which mimic the experiment evolution. 
A basic characteristic of the experiment is that it is completely 
irreversible, so the dynamics has to include this important feature.
We consider a discrete time evolution and the discrete dynamical
rule \cite{Duemmer2007} is given by:
\begin{equation}
\label{dynamics}
 h_i(t+1)-h_i(t)=v_i(t)= \theta \left[  f_i \right ] \;\;\;\; t=0,1,2 \ldots
\end{equation}
where $\theta$ is the Heaviside step function.
Then we apply the following procedure:
\begin{enumerate}
 \item Start at $t=0$ with a flat line located at $h=0$ setting $h_i=0 \;\; \forall i$.
 \item Compute the local force ($f_i$) at each site and using the dynamical
       rule (Eq. \ref{dynamics}) compute the local velocity of each site. 
       We can define the velocity of the string for this time, as $v(t)=\frac{1}{L} 
       \sum_{i=1}^{L} v_i(i)$.
 \item Advance the sites according their local velocities $v_i$.
 \item Generate new random forces for those sites that have been advanced.
 \item Go to step (2) and advance the \emph{simulation time} by one unit.
\end{enumerate}
This evolution shows a depinning transition at  $f_c \sim
1.88$ in which the velocity of the line $v(t \rightarrow \infty) > 0$ when $f>f_c$ 
and $v(t \rightarrow \infty )=0$ when $f<f_c$.

In order to simulate the creep evolution of the string we use an
external force below the depinning threshold, and when the line 
gets stuck we let thermal fluctuations play a role. 
We scan all the sites and set $v_i(t)=1$ with a probability $p=\exp \left (
\frac{f_i}{T_p} \right )$  and $v_i(t)=0$ with a probability $1-p$,
where $T_p$ is proportional to temperature. This can trigger an
avalanche which will have a finite duration $T$ since the system is 
below the depinning treshold. We define the avalanche size as $S=\sum_{T} v(t)$. 
If we consider small enough temperatures compared to the typical internal forces, the
avalanche needs some time to be triggered, which is defined as the
waiting time $\tau$. We define this waiting time as the time between the end of an avalanche and the starting time of the next one.
\begin{figure}[htb]
\begin{center}
\epsfig{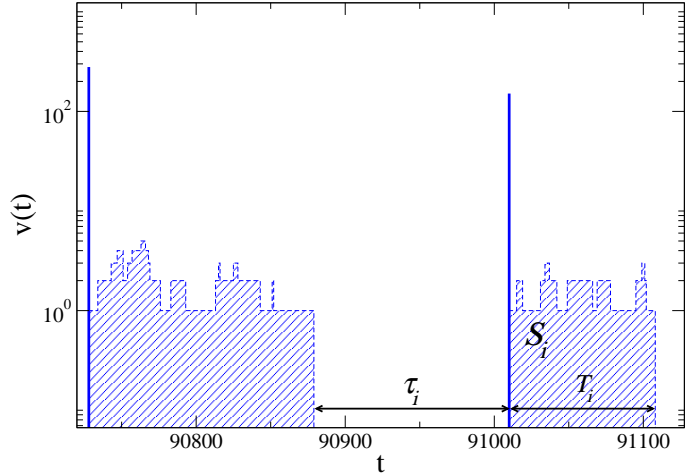}
\end{center}
\caption{\label{fig:sims_sign} Velocity of the long-range elastic string 
as a function of simulation time (dotted line).
The vertical and solid lines represents the signal $S(t)$ ploted in figure \ref{fig:et}(b).
Avalanche properties are also shown: $\tau_i$ is the waiting time, $T_i$ is the avalanche duration duration, and $S_i$ is the avalanche size. }
\end{figure}

In summary, this long-range elastic line model in the creep regime has
an avalanche-like behaviour.
Each avalanche is characterized by three quantities: Waiting time $\tau$, duration $T$, and size $S$ (see Fig.~ \ref{fig:sims_sign}).
Moreover, we observe that for long times, when the steady state is reached, durations are small compared to waiting times, for that reason we can simplify the signal just taking into account the starting time of the avalanche and its size.

\section{Creep velocity}

\begin{figure}[htb]
\begin{center}
\epsfig{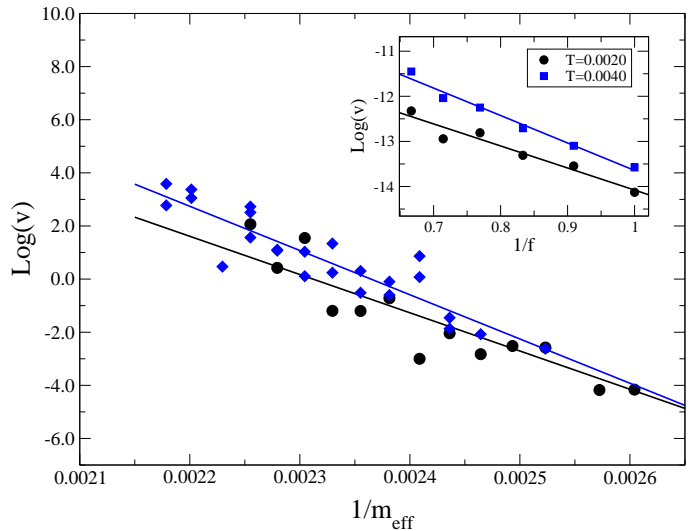}
\end{center}
\caption{\label{fig:speed_force}The creep velocity vs. the inverse
of the applied force or mass, $m_{eff} = m$. Inset: creep
velocity vs. $f$ for the simulation model for two different temperatures.}
\end{figure}

The main data about both simulations and experiment on the creep
velocity are shown in Figure \ref{fig:speed_force}. The prediction
of Eq. (\ref{vcreep}) is that the velocity is exponential in the
effective driving force. In the case of the experiments at hand, we
face the problem that we do not know the average fracture toughness 
$\langle K\rangle$ empirically. It depends on the loading geometry, and
on the material at hand. There are estimates for similar papers in
the literature in the mode I case (see e.g. Ref. \cite{karenlampi})
which indicate that the value of $\langle K\rangle$  is
lower by at least a factor of two compared with the actually used
loads. One can try to work around the problem by guessing 
$\langle K\rangle$ and checking how that affects the apparent
functional relationship of $v$ vs. the reduced mass $m_{eff} =m - \langle K\rangle $.
In the range of physically sensible values of $\langle K\rangle$ 
the velocity exponential behaviour does not change and thus we can take $m_{eff} =m$ .

From the figure we may conclude that the effective creep exponent
$\mu\sim1$, though there is variability among the data sets. One of
the data sets (black circles) shows some slight curvature. The main
finding, interpreted via Eq. (\ref{expo}) then indicates that the
effective roughness exponent $\zeta \sim 1/3$, which is the expected 
{\em equilibrium} value for a long-range elastic problem with $\alpha=1$
\cite{koivisto}.

The numerical simulation data agree qualitatively
with the exponential decay except very close to the depinning transition.
According to the creep formula (see e.g. \cite{kolton}), 
we should expect that the velocity of the long-range string was 
\begin{equation}
v(f,T_p)  \sim exp \left[  -\frac{C}{T_p} \left (\frac{1}{f} \right) ^\mu  \right ]. 
\end{equation}
However, it appears that slope as a function of the temperature is not exactly the expected one.
One reason is that the model is simplified: we only let thermal fluctuations 
act when the string gets stuck so avalanche nucleation during an avalanche is
neglected. This may be of importance very close to $f_c$ and for long avalanches.
\begin{figure}[htb]
\begin{center}
\epsfig{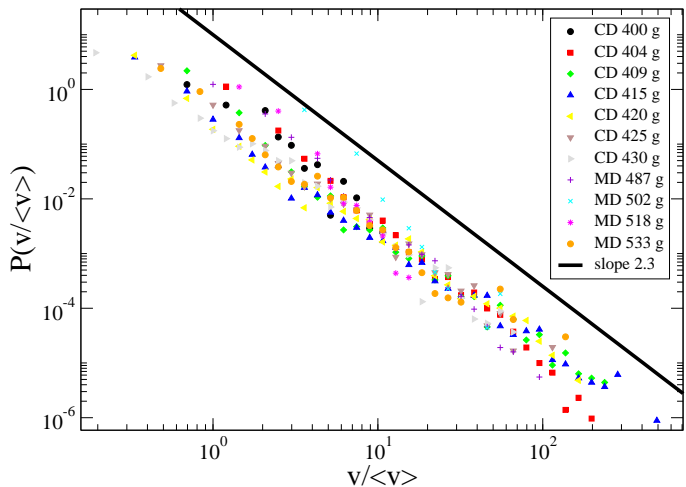}
\end{center}
\caption{\label{fig:stick}  Histogram of a normalized velocity
obtained from discretized distance data. Velocity, $v$, is an
average in a 0.5 s time window. $<v>$ is an average over experiments
with same weights.}
\end{figure}

The exponential average creep velocity can most directly be compared
with the measured velocities from the distance sensor over short time-spans. 
Figure \ref{fig:stick} shows the probability distributions $P(v)$ for a
very large number of different experiments, for the $v = \Delta h /
\Delta t$ with $\Delta t =$ 0.5 $s$. The general trend shows clear
stick-slip characteristics in the sense that the local
velocities vary with a power-law -like fashion. The typical slope of
the data is about -2.3 though a more detailed look indicates that
there is a tendency for the exponent to change with $m$ and with
$\Delta t$ (increasing both decreases the slope). It is an
interesting question of how this locally time-averaged velocity is
related to the average creep velocity, and the avalanches that
contribute to it, somewhat hindered by the relative large fluctuations
in the distance sensor - for which reason we resort in the detailed
avalanche dynamics studies to the AE.

\section{Statistical distributions}

Next we consider the statistics of the AE timeseries from the
experiments as signatures of the intermittent avalanche activity in
the system during creep. In our setup, we face the problem that
direct imaging of the front dynamics is if not impossible then
difficult to realize. Thus we take the AE data up to be scrutinized
as detailed information. It can be studied from the viewpoint of the
correlations of the creep avalanche activity but the finer
details thereof are left to the next Section.
Here, we consider the averaged distributions of three quantities: 
i) AE energy $E$ in experiments and  avalanche size in simulations 
   assuming that the avalanche energy is proportional to its size $E\sim S$
ii) waiting times $\tau$ both for  experiments and simulations 
and iii) avalanche durations $T$ only for simulations, because  
experimental avalanches have very short durations and
can be neglected. 
\begin{figure}[htb]
\begin{center}
\epsfig{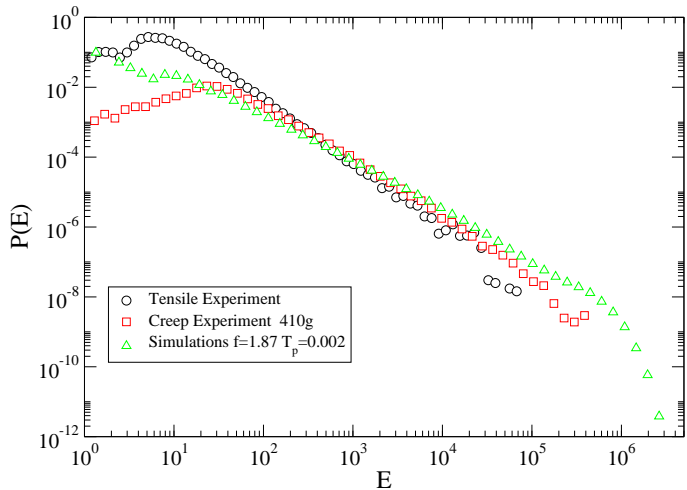}
\end{center}
\caption{\label{fig:PE} Energy distributions for the tensile experiment (circle), for the 
creep experiment (square), and for the simulations (triangle up). For the simulations we 
are ploting the histogram of the avalanches sizes $\{S_i\}$. We can consider that
the energy of an avalanche is proportial to its size, so $S_i \sim E_i\}$.}
\end{figure}
\begin{figure}[htb]
\begin{center}
\epsfig{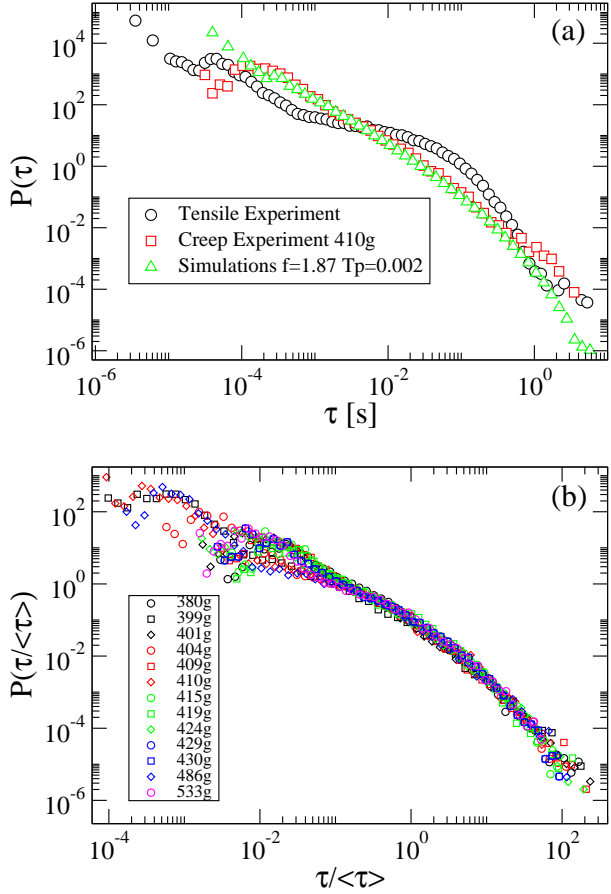}
\end{center}
\caption{\label{fig:Ptau} (a) Waiting time distributions for the tensile experiment (circle), 
for creep experiment with 410g (squares), and for the simulations with $f=1.87$ and $T=0.0020$ (triangles up). (b) Normalized waiting times for different creep experiments.}
\end{figure}

Figure \ref{fig:PE} shows three cases of the avalanche size
distributions. We compare the creep data for one mass $m$ 
to a similar dataset for a tensile
experiment done at a constant average front velocity \cite{epl}. 
Moreover data is included from the creep model for the parameters shown in the
caption. The normalization of the data for the experiments is such
that the $E_{min}$ has been scaled to unity. Recall that the events
are restricted in size from below by a thresholding applied to the
original AE amplitude signal $A(t)$, from which the events are
reconstructed. We can observe that the effective power-law
exponents of the experimental data are $\sim 1.6$ for the creep and 
$\sim 1.8 $ for the tensile cases, respectively. These are very close to each
other, while the simulation data results imply $\sim 1.4$ not very far
from the experimental values. We also can observe that there is no evident
cutoff in any of them (the bending in the simulations case is a finite
size effect). These data can be compared with the
Oslo plexiglass experiment where for the avalanche size
distribution the value of $\beta = 1.6\pm 0.1$ has been found \cite{maloy1,maloy2}.
\begin{figure}[htb]
\begin{center}
\epsfig{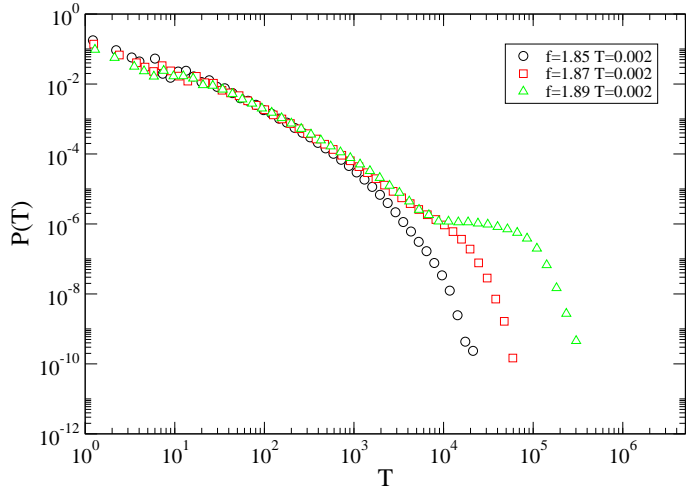}
\end{center}
\caption{\label{fig:PT} Avalanche duration distributions for $Tp=0.002$ and three different 
forces. For the case with $f=1.89$ we are above the depinning treshold.}
\end{figure}

The waiting times are reported in Fig. \ref{fig:Ptau}. For all the
three cases $P(\tau)$ is broad. In the tensile case, it is known
that there appears to be a "bump" in the distribution, or a typical
timescale. This is absent from the creep one. It is interesting to
note that here the simulation model agrees rather well with the
creep case. For larger $m$ it is possible that the waiting times
start to look more like the tensile case. We also present the scaled
distributions for all the experiments. Later, in the next section,
we discuss the attempt to link this to a background plus correlated,
triggered activity.

Finally, in Fig. \ref{fig:PT} we show the avalanche durations from
the simulations. In the case of the experiment this is more
complicated due to the fact that the actual amplitude signal is
convoluted via the preprocessing electronics and the response
function of the piezos with which the AE is measured. Later we
present some examples of the outcome, but here we just discuss the
clear-cut case of the simulations also since they give an idea about
what one might see in the experiment, ideally. The main points that
one learns from the Figure are that a true power-law-like $P(T)$
ensues only at the proximity of the $f_c$. For values higher or
lower than that the shape of the distribution changes, in particular
such that not only a cut-off appears but also the clear power-law
character starts to disappear.

\section{Measures of correlated dynamics}

\subsection{Correlations}

Next we look at the detailed temporal structure of the AE signal. 
The main question is whether the creep activity
exhibits interesting features that would in particular differ from
the theoretical expectations - based on elastic line depinning the
inter-avalanche correlations should be expected to be negligible.

\begin{figure}[htb]
\begin{center}
\epsfig{file=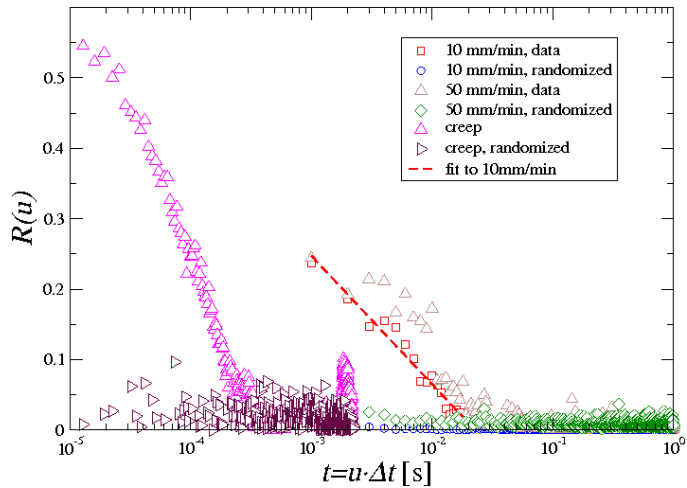,width=9cm,clip=}
\end{center}
\caption{\label{fig:creep-correlations} The autocorrelation function
of the averaged event energy in paper peeling under creep and
tensile loading modes. Comparisons to the randomized data are also
included. The numerical data and the corresponding randomized data
are not distinguishable.}
\end{figure}
In Figure \ref{fig:creep-correlations} we show the autocorrelation
function $R(t)$ of the event energy time series. The autocorrelation 
function is defined as:
\begin{equation}
  R(u) = \frac{\frac{1}{N}\sum_{t=1}^{N} E_t E_{t+u} -\langle E \rangle^2}
                                       {\langle E^2 \rangle -\langle E \rangle }
\end{equation}
where $E_t$ is the energy of the AE signal at time $t$ and $\langle E \rangle $ is the
average value of the energy. $E_t$ is defined as a sum of squared
amplitudes of the AE signal in the time interval $[t, t+\Delta
t]$. The length of the interval $\Delta t$ is chosen to be $10^{-3}s$
in the tensile and $10^{-5}s$ in the creep peeling experiment in
order to capture the correlations in both cases.

When compared to paper peeling experiments under a constant
strainrate, the correlation decays at a much faster rate than in
creep peeling experiments. In that case, the existence of a slow
decay might be taken to be connected to the fact that there is a
typical scale in the waiting time distribution which is not the case
for creep, seemingly. The functional form of the shown case of a logarithmically
decreasing autocorrelation function is $R(t) = -0.3 - 0.08 ln(t)$. The
data are also compared to a randomized timeseries, and one can see
that the correlations disappear. For the simulated data the
autocorrelation function shows no difference to a randomized
signal. All in all these results imply that there are contrary to
theoretical models temporal correlations, albeit in creep on a very
short timescale.

\begin{figure}[htb]
\begin{center}
\epsfig{file=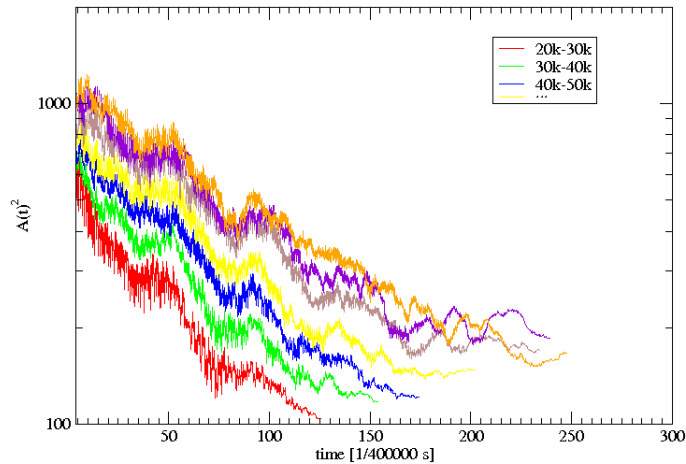,width=9cm,clip=}
\end{center}
\caption{ \label{fig:form-functions}  Squared amplitude of an event
averaged over all events in the creep peeling experiment. The
average is taken over events with different sizes and the size is
indicated as different colors in the figure.}
\end{figure}

In Figure \ref{fig:form-functions} we show an envelope event form
for different events with different event energies. We see an
exponential decay for the tail of the event, but the event envelope
becomes more extended in time when the energy of the event is
larger. Typical events extend up to 0.5ms, which corresponds to
decay of the correlation in the Fig. \ref{fig:creep-correlations}. A
correlation up to time-scales larger than the typical event length
is only seen in the strain-controlled peeling.

\begin{figure}[htb]
\begin{center}
\epsfig{file=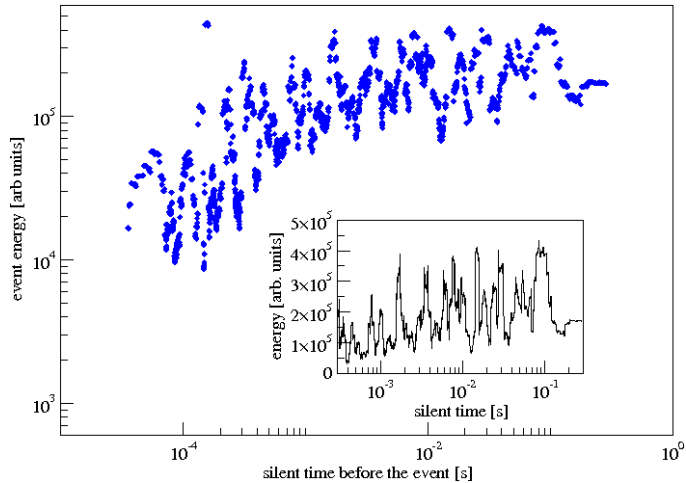,width=9cm,clip=}
\end{center}
\caption{\label{fig:silt} Averaged energy as a function of silent
time before the event with weight 410g.}
\end{figure}

Since the timeseries of AE is so intermittent it is better to
concentrate on measures that consider directly the avalanches. In
Figure \ref{fig:silt} we depict the averaged energy as a function of
a silent time before the event from paper peeling experiments in
creep. The event energy is in many datasets slightly correlated to the waiting time
before the event. This correlation disappears if one considers the
opposite case of the waiting time after the event. The suggested
interpretation is that the elastic fracture line apparently as a
physical system ages before a large event, while there is no real
dependence of the waiting time on the energy dissipated in the
previous event.

The difference in the autocorrelation between the creep and tensile
peeling experiments might be attributed to the forcing the line to
move in the latter, which induces a ``fiber-scale'' to results. This
is also supported by observing the waiting time distribution, where
the pdf deviates from a power-law.

In paper peeling we study the clustering of events by computing the
correlation integral $C(\Delta T)$, that is the probability that two events
are separated smaller time than $\Delta T$. The correlation integral is given
by:%
\begin{equation}
C(\Delta T) = \frac{2}{N(N-1)} \sum_{i<j} \theta\left[ \Delta T-(t_j-t_i) \right],
\label{eq:corint}
\end{equation}
where $N$ is number of events in the experiment and $t_i$ is the event
occurrence time. 

\begin{figure}[htb]
\begin{center}
\epsfig{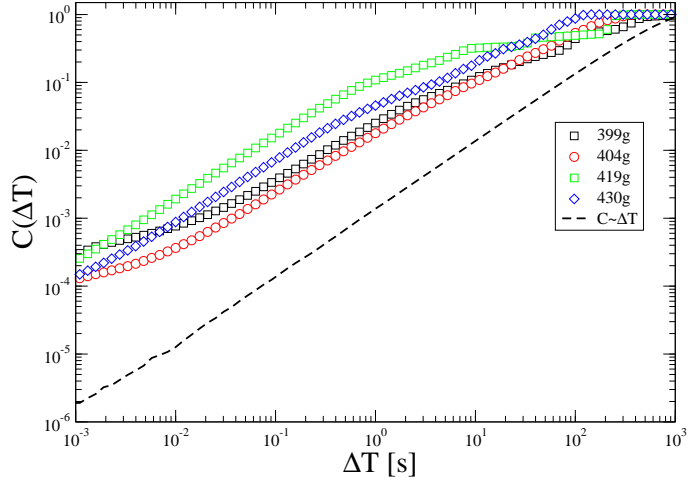}
\end{center}
\caption{\label{fig:corint-weight} Correlation integrals for the creep in peeling experiment.}
\end{figure}

Correlation integrals \cite{weiss} are shown in the Fig. \ref{fig:corint-weight}
for the peel creep experiment. If the probability of the event
occurrence is equal for every time interval, then one can assume that
correlation integral increases as $C(\Delta T) \sim \Delta T$. We see a power law
$\Delta T^{0.9}$ in sufficiently large times, but when the distance of events
approaches the experiment length we see small deflection in the
curve. At temporal scales of the order of $10^{-2}$s we see deviation from the
power law behaviour, which indicates event clustering.

\subsection{Seismicity - cascading occurrences as a model for the experimental data}

In this part we will show how fracture in heterogeneous material, such as line creep in paper peeling, behaves, in time, similarly to the rupture at the Earth scale, e.g. the earthquakes driven by plate tectonic deformation.

From seismology it is known that seismicity can be described by two processes: the background seismicity and the triggered events. 
The first one is modelled as a homogeneous Poisson process, while the second one as a power law decay of seismic rate following the occurrence of any event, e.g. the Omori's law 
\cite{Kagan,Utsu,Helmstetter_a}:
\begin{equation}
 \label{eq:sis1}
  R = \mu_0 + \sum_{t<t_i} \lambda_i(t)
\end{equation}

\noindent
The first term in the right hand side of Eq.~\ref{eq:sis1} is the background seismicity, while
the second term is the correlated part of the seismicity, that is, the superposition of 
time-dependent series of triggered seismicity following any event. 
The triggering process of the latter is reproduced by models of cascading effect for earthquake interactions, i.e. ETAS (Epidemic Type Aftershock Sequence) model \cite{Kagan,Utsu,Helmstetter_b}.
This stochastic point process is based on the Gutenberg-Richter law for energy distribution and Omori's
law for time distribution of seismicity rate. According to this model, the rate of aftershocks
triggered by an earthquake occurring at time $t_i$ with magnitude $M_i$ is given by:
\begin{equation}
\label{eq:sis2}
\lambda_i(t) = \frac{K_0}{(c+t-t_i)^p}10^{\alpha(M_i-mc)},
\end{equation}
where $K_0$, $\alpha$, $c$ and $p$ are constants and $mc$ is the completeness magnitude of the catalogue.

The total earthquake rate of Equation (\ref{eq:sis1}) is therefore the sum of all preceding earthquakes (triggered directly by the background events or indirectly by previous 
triggered events) and the constant background rate $\mu_0$. 
This model reproduces most of the statistical properties of earthquakes, including aftershock and foreshocks distributions in time, space and energy \cite{Helmstetter_a}.

\begin{figure}[htb]
\begin{center}
\epsfig{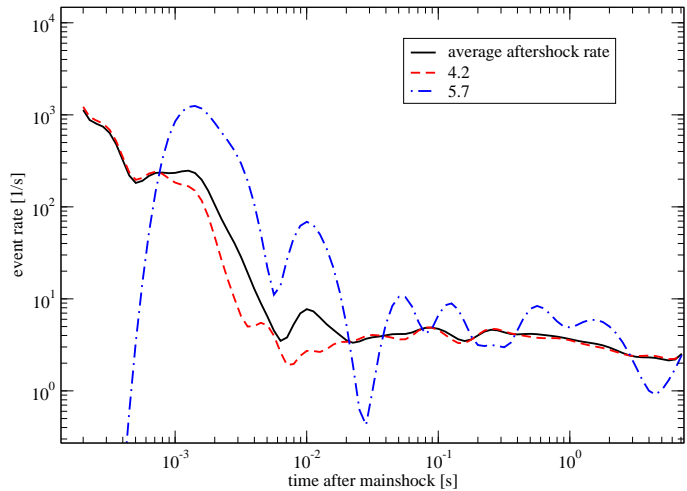}
\end{center}
\caption{\label{fig:sis1}
Event rate following events in paper peel creep experiments with 
$m=409g$. Time $t = 0$ is the target event occurrence. Aftershock rates are averaged within each magnitude class of target event (blue line: $4.2 - 5.72$;
red line $5.72 - 7.2$). We compute the magnitude class $M=log_{10} (E_{M})$ where $E_{M}$ is the energy of the target event.
All magnitude classes are averaged together (thick black line). Correlation between events is characterized by a power law decay of the activity after the target event. The time for which events are correlated is a function of the target event magnitude, as well as the number of triggered events (see Equations (\ref{eq:sis1}) and (\ref{eq:sis2})). The observed duration of the aftershock sequence is bounded by the level of the background uncorrelated constant rate.}
\end{figure}

Figure \ref{fig:sis1} illustrates the average acoustic event rate following any event for the peel creep experiments (load $m= 409 g$). 
It is reminiscent of Omori's law for tectonic seismicity, where we can observe the power law decay representing the cascade of aftershocks following an event. For times greater than $10^{-2}$ seconds,
the event rate keeps constant, at the background rate level. The exponent of the power law decay of
event rate is equal to $1.5 \pm 0.1$. This is fairly close to what would fit the experimental $P(\tau)$ demonstrated in Fig. \ref{fig:Ptau}.

The AE, triggered by line creep in paper peeling, is characterized by power
law distribution on energy (Fig. ~\ref{fig:PE}) and power law relaxation of aftershock rate (Fig. ~\ref{fig:sis1}). ETAS style models reproduce these macroscopic patterns, including foreshocks as aftershocks of conditional mainshocks \cite{Helmstetter_b}.
Corral \cite{Corral} shows that the inter-event time probability density for such kind of ETAS model for event occurrences follows a gamma distribution, according to: 
\begin{equation}
\label{eq:sis3}
p(\tau) = C \tau^{\gamma-1} \exp({-\tau/\beta})
\end{equation}
where $\tau$ is the normalized inter-event time obtained by multiplying the inter-event
time $\delta t$ with the earthquake rate $\lambda$, that is $\tau = \delta t \lambda$. 

\begin{figure}[htb]
\begin{center}
\epsfig{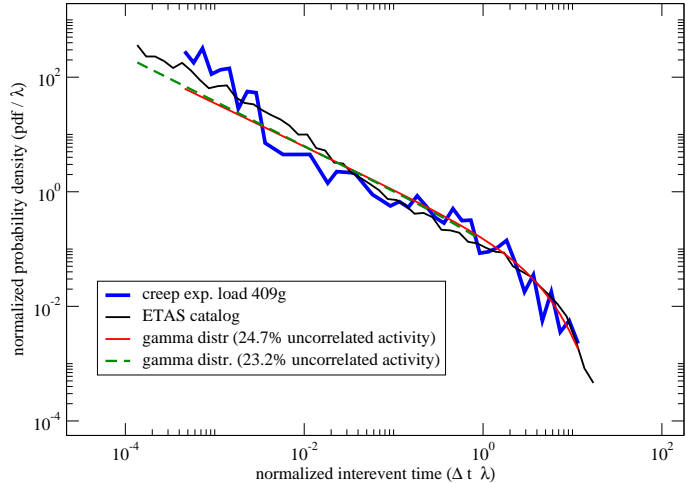}
\end{center}
\caption{\label{fig:sis2} Inter-event time probability distribution for experimental 
dataset (thin red curve) and synthetic catalogue generated by ETAS model (thin black curve). Dotted thick curves are gamma distribution fits to data and ETAS model (red dotted line for the real data and black dotted curve for ETAS). Estimations of background fraction of events according to Hainzl et al. (2006) technique are close together ($23-25\%$) for both data and simulation.
ETAS parameters are: $p = 1.4$, $K_o = 0.09$, $\alpha = 0.9$ ($n = 0.9$), $b = 1$, $c=0.001s$.}
\end{figure}

Molchan \cite{molchan} showed that, in agreement with Equation (\ref{eq:sis3}), the distribution decays exponentially for large inter-event times and that the value 
$1/\beta$ is the fraction of mainshocks among all seismic events. 
According to Hainzl et al. \cite{Hainzl}, $1/\beta$ is a regional quantity, 
allowing for non-parametric estimate of the background rate in a specific process.
In order to simulate the AE properties of the creep fracture experiment ($m=409g$), we tuned an ETAS model to fit the estimated percentage of background activity of real
data. One must notice that robust inversion of ETAS model parameters is not yet
available.  Figure \ref{fig:sis2} shows the comparison between inter-event time distributions of a synthetic catalogue generated by ETAS model. Both, simulations and data inter-event time distributions fit a gamma distribution. Other possible data fittings are possible \cite{Saichev}, but this lies outside our aim of comparison between data from paper peeling and ETAS simulations. The fit may underestimate here (see also Fig. \ref{fig:Ptau}b) slightly the exponent of the
power-law part of the waiting-time distribution. In any case, the relevant exponent here is definately smaller than in the case of rock fracture \cite{davidsen} ($p=1.4$).

To summarize, line creep in paper peeling at a scale of $\sim 10^{-1}m$ and 
$\sim 10^2s$ triggers brittle creep damage that seems to share the same generic temporal properties than the ones observed for tectonic seismicity at scales of 
$\sim 10^6 m$, $\sim 10^2$ years. 
These properties can be reduced to a rough constant seismicity rate with bursts of 
correlated activity, contemporary to power law distribution of event 
sizes and (short-time) inter-event times.
Estimates of Omori's law exponent suggest a faster relaxation for the paper peeling 
case than for Earth crust response to tectonic loading, p equal to 1.4 and 1 
respectively \cite{Utsu}. The portion of uncorrelated events suggests a 
slightly lower triggered event rate in paper peeling than in the Earth crust 
deformation. Estimations of the background portion of AE did not show any sensitive
dependence on the applied loading. Whether the difference between paper experiments and earthquakes come from experimental conditions or fracturing mode (i.e. tensile, creep or compression) remains an open question. For earthquakes no change in relative portions of background and triggered activity is resolved for compression, extensional or shear tectonic settings.

\section{Conclusions}

We have overviewed a simple creep experiment which uses paper and can
be studied to investigate planar crack propagation in a disordered
medium. The information that one can obtain and then compare to
relevant theory extends from the average front velocity to details of
the spatiotemporal dynamics. We have also for a comparison studied a
classical non-local elastic line model under creep conditions. This
shows similar features to the experiment: an exponential dependence of
the creep velocity on the applied force or mass or stress-intensity
factor.

The typical statistical distributions are power-law -like in
particular for the event energy/size. It is perhaps useful to recall
that the waiting time distribution is quite broad. There is
currently no understanding as to why, in particular one should note
that the current experimental setup allows to study this issue in a
steady-state unlike in most other fracture-related creep tests. In
general as such distributions are regarded the line creep model
agrees at least qualitatively with the experimental data. Our
results are also in line with other similar planar crack data
(though these are obtained usually in the constant-velocity
ensemble, not in creep \cite{maloy1,maloy2,bonamy}).

Looking in more detail at the correlations of the activity,
differences transpire however. The experimental AE events show
subtle correlations via the autocorrelation function, via the
waiting times before events, and via the Omori's law. All these
measure different aspects of the avalanche activity, and in all the
cases the model differs in its behavior. Here, we lack completely
theoretical understanding, in particular as regards such a
quantitative measure as the Omori exponent. It is interesting to note that 
geophysics -oriented analysis methods produce results in agreement with
observations from tectonic activity. Here again the steady-state character of
the experiment at hand is of utility.

In the future such experiments and such comparisons can be used to
study several different aspects of avalanching systems, creep
fracture, and models for line depinning. A particularly pertinent
question is for instance whether rate-dependent processes in the
material at hand modify the kinetics of the creep in some suitable
way that still maintains the creep vs. force -relation intact.
We shall ourselves attempt a more careful study of the creep
model, and analyze how its correlation patterns could be matched
with the experiment.

 {\it Acknowledgements -} The authors would like to
thank for the support of the Center of Excellence -program of the
Academy of Finland, and the financial support of the European
Commissions NEST Pathfinder programme TRIGS under contract
NEST-2005-PATH-COM-043386. MJA is grateful for the hospitality of
the Kavli Institute of Theoretical Physics, China in Beijing, where
the work at hand was to a large degree completed. Discussions with
Lasse Laurson (Helsinki), St\'ephane Santucci (Oslo), 
Daniel Bonamy (Saclay), and Stefano Zapperi (Modena) are also acknowledged.


\begin{thebibliography}{99}

\bibitem{advphys}
M.\ J.\ Alava, P.\ K.\ V.\ V.\ Nukala, and S.\ Zapperi, Adv.\ Phys.\ {\bf 55},
349 (2006).

\bibitem{sethrev}
J.\ P.\ Sethna, K.\ A.\ Dahmen, and C.\ R.\ Myers, Nature {\bf 410}, 242 (2001).

\bibitem{Kertesz}
 J.\ Kert{\'e}sz, V.\ K.\ Horv{\'a}th, and F.\ Weber, Fractals {\bf 1}, 67 (1993).

\bibitem{oma}
L.\ I.\ Salminen, A.\ I.\ Tolvanen, and M.\ J.\ Alava,
Phys. Rev. Lett. {\bf 89}, 185503 (2002).

\bibitem{santucci} 
S.\ Santucci, L.\ Vanel, and S.\ Ciliberto, Phys.\ Rev.\ Lett.\ {\bf 93}, 095505 (2004).

\bibitem{epl}
L.\ I.\ Salminen, J.\ M.\ Pulakka, J.\ Rosti, M.\ J.\ Alava, and K.\ J.\ Niskanen, Europhys.\ Lett.\ {\bf 73}, 55 (2006).

\bibitem{koivisto}
J.\ Koivisto, J.\ Rosti, and M.\ J.\ Alava,
Phys.\ Rev.\ Lett.\ {\bf 99}, 145504 (2007).

\bibitem{Rosso2002}
A.\ Rosso and W.\ Krauth, Phys.\ Rev.\ E {\bf 65}, 025101(R) (2002).

\bibitem{maloy1}
J.\ Schmittbuhl and K.\ J.\ M{\aa}l{\o}y, Phys.\ Rev.\ Lett.\ {\bf 78}, 3888 (1997).

\bibitem{maloy2}
K.\ J.\ M{\aa}l{\o}y, S.\ Santucci, J.\ Schmittbuhl, and R.\ Toussaint,
Phys.\ Rev.\ Lett.\ {\bf 96}, 045501 (2006).

\bibitem{Fisher}
D.\ S.\ Fisher, Phys.\ Rep.\  {\bf 301}, 113 (1998).

\bibitem{Fisher2} 
S.\ Ramanathan and D.\ S.\ Fisher, Phys.\ Rev.\ Lett.\ {\bf 79}, 877 (1997).

\bibitem{schmitt} 
J.\ Schmittbuhl, S.\ Roux, J.\ P.\ Vilotte, and K.\ J.\ M{\aa}l{\o}y, 
Phys.\ Rev.\ Lett.\ {\bf 74}, 1787 (1995).

\bibitem{krauth}
A.\ Rosso and W.\ Krauth, Phys.\ Rev.\ Lett.\ {\bf 87}, 187002 (2001).

\bibitem{creep}
T.\ Nattermann, Europhys.\ Lett.\ {\bf 4}, 1241 (1987);
L.\ B.\ Ioffe and V.\ M.\ Vinokur, J.\ Phys.\ C {\bf 20}, 6149 (1987);
T.\ Nattermann, Y.\ Shapir, and I.\ Vilfan, Phys.\ Rev.\ B {\bf 42}, 8577 (1990).

\bibitem{chauve}
P.\ Chauve, T.\ Giamarchi, and P.\ Le Doussal, 
Phys.\ Rev.\ B {\bf 62}, 6241 (2000).

\bibitem{kolton}
A.\ B.\ Kolton, A.\ Rosso, T.\ Giamarchi, and W.\ Krauth, 
Phys.\ Rev.\ Lett.\ {\bf 94}, 047002 (2005).

\bibitem{lemerle}
S.\ Lemerle, J.\ Ferr\'e, C.\ Chappert, V.\ Mathet, T.\ Giamarchi, and P.\ Le Doussal,
Phys.\ Rev.\ Lett.\ {\bf 80}, 849 (1998).

\bibitem{crex}
Th.\ Braun, W.\ Kleemann, J.\ Dec, and P.\ A.\ Thomas,
Phys.\ Rev.\ Lett.\ {\bf 94}, 117601 (2005); 
T.\ Tybell, P.\ Paruch, T.\ Giamarchi, and J.\-M.\ Triscone, 
Phys.\ Rev.\ Lett.\ {\bf 89}, 097601 (2002).

\bibitem{Alava2006}
M.\ J.\ Alava and K.\ Niskanen, Rep.\ Prog.\ Phys {\bf 69}, 669 (2006)

\bibitem{Tanguy1998}
A.\ Tanguy, M.\ Gounelle, and S.\ Roux, Phys.\ Rev.\ E {\bf 58}, 1577 (1998).

\bibitem{Duemmer2007}
O.\ Duemmer and W.\ Krauth, J.\ Stat.\ Mech. {\bf P01019} (2007).

\bibitem{karenlampi} 
Y.\ Yu and P.\ K\"arenlampi, J.\ Mat.\ Sci.\ {\bf 32}, 6513 (1997).

\bibitem{weiss}
J.\ Weiss and D.\ Marsan, Science {\bf 299}, 89 (2003).

\bibitem{Kagan}
Y.\ Y.\ Kagan and L.\ Knopoff, J.\ Geophys.\ Res.\ {\bf 86} (B4), 2853 (1981).

\bibitem{Utsu}
T.\ Utsu, Y.\ Ogata, and S.\ Matsuura, J.\ Phys.\ Earth {\bf 43}, 1 (1995).

\bibitem{Helmstetter_a}
A.\ Helmstetter, D.\ Sornette, Phys.\ Rev.\ E {\bf 66}, 061104 (2002).

\bibitem{Helmstetter_b}
A.\ Helmstetter, D.\ Sornette and J.\-R.\ Grasso, 
J.\ Geophys.\ Res.\ {\bf 108} (B1), 2046 (2003).

\bibitem{Corral}
A.\ Corral, Phys.\ Rev.\ Lett.\ {\bf 92}, 108501 (2004).

\bibitem{molchan}
G.\ Molchan, Pure Appl.\ Geophys.\ {\bf 162} 1135 (2005).

\bibitem{Hainzl}
S.\ Hainzl and Y.\ Ogata, J.\ Geophys.\ Res.\ {\bf 110} (B05), S07 (2002).

\bibitem{Saichev}
A.\ Saichev and D.\ Sornette, J.\ Geophys.\ Res.\ {\bf 112} (B04), 313 (2007).

\bibitem{davidsen}
J.\ Davidsen, S.\ Stanchits, and G.\ Dresen, Phys.\ Rev.\ Lett.\ {\bf 98}, 125502 (2007).

\bibitem{bonamy}
D.\ Bonamy, L.\ Ponson, S.\ Prades, E.\ Bouchaud, and C.\ Guillot,
Phys.\ Rev.\ Lett.\ {\bf 97}, 135504 (2006).

\end{thebibliography}
\end{document}